\documentclass[aps,prb,10pt,twocolumn,amsmath,amssymb,floatfix,superscriptaddress]{revtex4-1}

\usepackage{graphicx}
\usepackage{rotating}
\usepackage{xspace}
\usepackage{tikz}
\usepackage{threeparttable}
\usepackage{geometry}
\usepackage{booktabs}
\usepackage{comment}
\usepackage{longtable}
\usepackage[normalem]{ulem}

\usepackage{hyperref}
\usepackage{afterpage}
\usepackage{siunitx}

\definecolor{robgreen}{rgb}{0.0, 0.425, 0.0}

\newcommand{\be}{\begin{enumerate}}
\newcommand{\ee}{\end{enumerate}}
\newcounter{saveenumi}

\newcommand{\ins}[1]{\textcolor{brown}{[ins:#1]}}

\newcommand{\dash}{\multicolumn{1}{c}{-}}

\newcommand{\mto}{MgTi$_2$O$_4$\xspace}
\newcommand{\cis}{CuIr$_2$S$_4$\xspace}
\newcommand{\cise}{CuIr$_2$Se$_4$\xspace}

\newcommand{\rAA}{\AA$^{-1}$\xspace}
\newcommand{\ttg}{$t_{2g}$\xspace}

\newcommand{\dd}{\mathrm{d}}

\newcommand{\nba}[1]{}

\newcommand{\qmax}{\ensuremath{Q_{\mathrm{max}}}\xspace}
\newcommand{\qmin}{\ensuremath{Q_{\mathrm{min}}}\xspace}

\newcommand{\pdfgui}{\textsc{PDFgui}\xspace}

\newcommand{\pdfgetxthree}{\textsc{PDFgetX3}\xspace}

\newcommand{\pyfai}{\textsc{pyFAI}\xspace}
\newcommand{\xpdf}{\textsc{xPDFsuite}\xspace}
\newcommand{\cmi}{\textsc{DiffPy-CMI}\xspace}

\renewcommand{\sout}[1]{}
\renewcommand{\ins}[1]{#1}

\begin{document}

\title{Two-orbital degeneracy lifted state as a local precursor to a metal-insulator transition}

\author{Long~Yang}
\affiliation{Department of Applied Physics and Applied Mathematics, Columbia University, New York, NY 10027, USA}

\author{Robert~J. Koch}
\affiliation{Condensed Matter Physics and Materials Science Division, Brookhaven National Laboratory, Upton, NY 11973, USA}

\author{Hong~Zheng}
\affiliation{Materials Science Division, Argonne National Laboratory, Argonne, IL 60439, USA}

\author{J.~F.~Mitchell}
\affiliation{Materials Science Division, Argonne National Laboratory, Argonne, IL 60439, USA}

\author{Weiguo~Yin}
\affiliation{Condensed Matter Physics and Materials Science Division, Brookhaven National Laboratory, Upton, NY 11973, USA}

\author{Matthew~G.~Tucker}
\affiliation{Neutron Scattering Division, Oak Ridge National Laboratory, Oak Ridge, TN 37830, USA}

\author{Simon~J.~L.~Billinge}
\affiliation{Department of Applied Physics and Applied Mathematics, Columbia University, New York, NY 10027, USA}
\affiliation{Condensed Matter Physics and Materials Science Division, Brookhaven National Laboratory, Upton, NY 11973, USA}

\author{Emil~S.~{Bozin}}
\email{bozin@bnl.gov}
\affiliation{Condensed Matter Physics and Materials Science Division, Brookhaven National Laboratory, Upton, NY 11973, USA}

\date{\today}

\begin{abstract}
The recent discovery of a local fluctuating $t_{2g}$ orbital-degeneracy-lifted (ODL) state in \cis as a high temperature precursor to the metal-insulator transition (MIT) opens the door to a possible widespread presence of precursor states in scarcely studied high-temperature regimes of transition-metal-based quantum materials.
Although in \cis the ODL state comprises one orbital per Ir, there is no fundamental reason to exclude multi-orbital ODL states in general.
The transition-metal spinel \mto exhibits a MIT on cooling below T$_{s}\approx$ 250~K, accompanied by long-range Ti $t_{2g}$ orbital ordering and spin singlet dimerization with associated average symmetry reduction to tetragonal.
It shares with \cis the pyrochlore transition metal sublattice with active $t_{2g}$ orbitals.
This, together with its different orbital filling ($t_{2g}^{1}$ versus $t_{2g}^{5.5}$) make it a candidate for hosting a multi-orbital ODL precursor state.
By combining x-ray and neutron pair distribution function (PDF) analyses to track the evolution of the local atomic structure across the MIT we find that {\it local} tetragonality already exists, in the metallic {\it globally} cubic phase at high temperature.
Local distortions are observed up to at least 500~K, the highest temperature assessed in this study.
Significantly, the high temperature local state revealed by PDF is not continuously connected to the orbitally ordered band insulator ground state, and so the transition cannot be characterized as a trivial order-disorder type.
The shortest Ti-Ti bond lengths corresponding to spin singlet dimers expand abruptly on warming across the transition, but they are still shorter than those seen in the cubic average structure.
These seemingly contradictory observations can be understood within the model of a local fluctuating two-orbital $t_{2g}$ ODL precursor state.
The ODL state in \mto has a correlation length of about 1~nm at high temperature.
We discuss that this extended character of the local distortions is consistent with the two-orbital nature of the ODL state imposed by the charge filling and the bond charge repulsion.
\end{abstract}

\maketitle

\section{Introduction}

The AB$_2$X$_4$ transition-metal spinels, with B cations forming a geometrically frustrated pyrochlore lattice~\cite{SickafusStructureSpinel1999}, exhibit a wealth of interesting electronic and magnetic properties, such as spin dimerization~\cite{RadaelliFormationisomorphicIr32002b,SchmidtSpinSingletFormation2004}, spin-lattice instability~\cite{MatsudaSpinlatticeinstability2007}, orbital ordering~\cite{RadaelliOrbitalorderingtransitionmetal2005a}, charge ordering~\cite{WrightLongRangeCharge2001}, and metal-insulator transitions~\cite{NagataMetalinsulatortransitionspineltype1998b,IsobeObservationPhaseTransition2002,ItoPressureInducedSuperconductorInsulatorTransition2003}.
For instance, the \cis thiospinel system exhibits a transition from paramagnetic metal to diamagnetic insulator on cooling~\cite{FurubayashiStructuralMagneticStudies1994c,MatsunoPhotoemissionstudymetalinsulator1997b,MatsumotoMetalinsulatortransitionsuperconductivity1999a}, the formation of structural isomorphic octamers~\cite{IshibashiXraydiffractionstudy2001a,RadaelliFormationisomorphicIr32002b}, and anomalous electrical properties~\cite{BurkovAnomalousresistivitythermopower2000b,TakuboIngapstateeffect2008b}.
In particular, \cis has broken symmetries in its ground state accompanied by formation of charge order, orbital order, and the creation of magnetic spin singlets on dimerized Ir$^{4+}$-Ir$^{4+}$ pairs~\cite{RadaelliFormationisomorphicIr32002b}.
These dimers were shown to disappear on warming through the transition, but preformed local symmetry broken states were seen at temperatures well above the global long-range ordering structure transition~\cite{BozinLocalorbitaldegeneracy2019c}.
Thus, on cooling, a precursor state exists that has broken local symmetry but is high symmetry over long range.
This behavior was shown to be driven by breaking of $d$-electron orbital degeneracies, resulting in a local fluctuating orbital-degeneracy-lifted (ODL) state distinct from spin singlet dimer.
This local state results from direct \ttg Ir orbital overlap promoted by the topology of the crystal structure.
In the regime of partial filling and high crystal symmetry that imposes the degeneracy of the orbital manifold, a molecular-orbital-like state is formed, accompanied by local structure distortion.
Many of the interesting physical properties of \cis could be explained by the short-range-ordered ODL mechanism, such as the non-conventional conduction in the high-temperature metallic phase~\cite{BurkovAnomalousresistivitythermopower2000b}, and the apparently contradictory observation of the destruction of the dimers at the phase transition but the persistence of poor metallic response above the transition~\cite{BozinDetailedMappingLocal2011c}.

Similar local symmetry-broken ODL states preformed at high temperature have been observed in other non-spinel $d$-electron systems such as the FeSe superconductor~\cite{KochRoomtemperaturelocal2019d,FrandsenQuantitativecharacterizationshortrange2019a}.
Even the well studied physics of the perovskite lanthanum manganites (LaMnO$_{3}$) could be interpreted in the same way~\cite{qiu;prl05}, arguing that the ODL phenomenon may be widespread among the myriad of materials with partially filled $d$-electron manifolds.
To explore this hypothesis it is important to seek it out systematically and characterize other materials that have the potential of revealing both commonalities and novel aspects that give more insights into the general ODL phenomenon.
For instance, there is no fundamental reason for the ODL states to be exclusively comprised of one orbital per transition metal, yet multi-orbital ODL states have so far not been identified.
Here we show that such multi-orbital ODL state exists in a related spinel, \mto, and that this is a consequence of the $d^{1}$ electron configuration decorating the pyrochlore lattice.

\mto\ shares many features with \cis~\cite{KhomskiiOrbitallyInducedPeierls2005a}.
They are both cubic spinels at high temperature with a transition metal ion pyrochlore sublattice of corner-shared tetrahedra.
In both materials the crystal field splits the $d$-orbitals into a $t_{2g}$ triplet and an $e_g$ doublet with the $t_{2g}$ orbitals partially occupied and the $e_g$ orbitals empty.
They both exhibit a temperature-dependent metal-insulator transition (MIT)~\cite{NagataMetalinsulatortransitionthiospinel1994a,IsobeObservationPhaseTransition2002}, the origin of which has been attributed to an orbital-selective Peierls mechanism~\cite{KhomskiiOrbitallyInducedPeierls2005a}.
They both exhibit a global symmetry lowering, from cubic to tetragonal, at the MIT on cooling.
They both have anomalous electrical resistivity behavior in the high-temperature metallic phase~\cite{ZhouOpticalstudyMgTi2O42006,BurkovAnomalousresistivitythermopower2000b}.
The symmetry lowering at the MIT is accompanied by a dimerization of transition metal ions that results in alternating short and long metal-metal bonds, and a resulting tetramerization~\cite{CroftMetalinsulatortransitionmathrmCuIr2003a}, along linear chains of ions on the pyrochlore sublattice~\cite{KhomskiiOrbitallyInducedPeierls2005a}.
The short bonds are associated with spin singlet dimer formation~\cite{RadaelliFormationisomorphicIr32002b,SchmidtSpinSingletFormation2004}.
The charge filling is also electron-hole symmetric between the systems with Ti$^{3+}$ having one electron in the $t_{2g}$ manifold, whilst Ir$^{4+}$ has one hole, although the nominal charge of Ir in \cis is 3.5+ which would place half a hole per Ir in the $t_{2g}$ anti-bonding band on average in this compound.

Despite the similarities, there are also notable differences.
The Ti valence electrons reside in 3$d$ orbitals whereas for Ir they are 5$d$, which are more extended and should result in larger bandwidth.
Indeed, the average separation between the transition metals on the undistorted tetrahedral pyrochlore sublattices of their cubic structures follow this expectation: Ti-Ti separation is shorter ($\sim$3.0~\AA) than Ir-Ir separation ($\sim$3.5~\AA).
Also, experimentally, in \mto the tetragonal distortion is shown to be compressive ($c < a$) below the MIT~\cite{SchmidtSpinSingletFormation2004} while it is tensile ($c > a$)~\cite{FurubayashiStructuralMagneticStudies1994c} in \cis.
Dimers form helical superstructures in \mto, whereas they form octameric ``molecules"
 in \cis, thereby lowering the symmetry further to triclinic.
Both materials tetramerize below the MIT; however, \cis has a 3+-3+-4+-4+ charge ordering (CO) that accompanies an orbital ordering (OO)~\cite{BozinLocalorbitaldegeneracy2019c}, whereas a uniform 3+ charge on Ti rules out CO in \mto.
With both similarities -- $t_{2g}$ orbitals on pyrochlore lattice, formation of dimerized singlets -- and differences to \cis\ -- lack of charge order, electron rather than hole states -- \mto\ provides a natural next step in a deeper, broader mapping of ODL phenomena.
As we show below, this includes the emergence of a multi-orbital ODL state.


Neutron PDF (nPDF) analysis on \mto has been performed previously~\cite{TorigoeNanoscaleicetypestructural2018} to study the spin singlet dimers, suggesting that they do persist to high temperature. However, due to the weak and negative neutron scattering length of Ti, and appreciable overlap with substantially stronger oxygen signal, nPDF itself cannot fully reveal how the local structure behaves with temperature.
Here we have applied a combined x-ray and neutron analysis to understand the full picture.
We find unambiguously that the Ti-Ti dimers do disassemble on warming through MIT.   However, the local structure does not agree with the average structure even at high temperature.
In analogy with \cis,  partially filled \ttg transition metal orbital manifolds of \mto, which are triply degenerate in the average cubic symmetry, utilize their favorable overlaps fostered by the pyrochlore sublattice topology to form an ODL state. Its structural signatures are observed up to at least 500~K ($\sim$ 2T$_{s}$), the highest temperature measured.
The spatial extent of the local structural response is found to be greater than that observed in \cis, consistent with the proposed two-orbital character of the ODL state in \mto.

\section{Methods} \label{sec;methods}
{\it Sample preparation \& characterization.\textemdash}
TiO$_{2}$, Ti metal, and an excess of MgO were mixed and reacted using a spark plasma sintering technique in a graphite crucible.
Synthesis at 1100~$^{o}$~C was complete in $\approx$ 15 minutes.  The sample was reground and fired a second time under similar conditions.
Powder X-ray diffraction analysis of the product showed well-crystallized MgTi$_{2}$O$_{4}$ spinel accompanied by an extremely small concentration of Ti$_{2}$O$_{3}$ as a second phase.
Magnetization measurements were conducted using SQUID magnetometer on a specimen with mass of 6.2~mg.
The data show a pronounced low temperature Curie tail which was subtracted.
The Curie-Weiss fit to the low temperature data yielded a Weiss temperature of -0.45~K, consistent with isolated spins, and a Curie constant of 0.023 emu$\cdot$K/mol which corresponds to $\approx 3$~\%\ by mole of putative Ti$^{3+}$ spin-1/2 impurities

{\it The PDF method.\textemdash}The local structure was studied using the atomic pair distribution function (PDF) technique~\cite{egami;b;utbp12,billi;b;itoch18}. The PDF analysis of x-ray and neutron powder diffraction datasets has been demonstrated to be an excellent tool for revealing local-structural distortions in many systems~\cite{BozinLocalorbitaldegeneracy2019c,KochRoomtemperaturelocal2019d,billi;prl96,qiu;prl05,YoungApplicationspairdistribution2011a,Keencrystallographycorrelateddisorder2015a,LavedaStructurepropertyinsights2018}.
The PDF gives the scaled probability of finding two atoms in a material a distance $r$ apart and is related to the density of atom pairs in the material. It does not presume periodicity so goes well beyond just well ordered crystals~\cite{egami;b;utbp12,billi;b;itoch18}. The experimental PDF, denoted $G(r)$, is the truncated Fourier transform of the reduced total scattering structure function~\cite{farro;aca09}, $F(Q)=Q[S(Q)-1]$:
\begin{equation}
\label{eq:FTofSQtoGr}
  G(r) = \frac{2}{\pi}
          \int_{\qmin}^{\qmax} F(Q)\sin(Qr) \: \dd Q,
\end{equation}
where $Q$ is the magnitude of the scattering momentum transfer. The total scattering structure function,
$S(Q)$, is extracted from the Bragg and diffuse components of x-ray, neutron or
electron powder diffraction intensity.  For elastic scattering, $Q = 4 \pi
\sin(\theta) / \lambda$, where $\lambda$ is the wavelength of the probe and
$2\theta$ is the scattering angle. In practice, values of $\qmin$ and $\qmax$
are determined by the experimental setup and $\qmax$ is often reduced below the
experimental maximum to eliminate noisy data from the PDF since the signal to
noise ratio becomes unfavorable in the high-$Q$ region~\cite{egami;b;utbp12}.

{\it X-ray PDF experiment.\textemdash} The synchrotron x-ray total scattering measurements were carried out at the PDF beamline (28-ID-1) at the National Synchrotron Light Source II (NSLS-II) at Brookhaven National Laboratory (BNL) using the rapid acquisition PDF method (RAPDF)~\cite{chupa;jac03}. The \mto powder sample was loaded in a 1~mm diameter polymide capillary and measured from 90~K to 500~K on warming using a flowing nitrogen cryostream provided by Oxford Cryosystems 700 Series Cryocooler.
The experimental setup was calibrated by measuring the crystalline Ni as a standard material. A two-dimensional (2D) PerkinElmer area detector was mounted behind the samples perpendicular to the primary beam path with a sample-to-detector distance of 227.7466~mm. The incident x-ray wavelength was  0.1668~\AA.
The PDF instrument resolution effects are accounted for by two parameters in modeling, $Q_{damp}$ and $Q_{broad}$~\cite{proff;jac99,farro;jpcm07}. For x-ray PDF measurement, these were determined as $Q_{damp} = 0.039$~\AA$^{-1}$ and $Q_{broad} = 0.010$~\AA$^{-1}$ by fitting the x-ray PDF from a well crystallized sample of Ni collected under the same experimental conditions.

In order to verify the data reproducibility, an additional set of x-ray data (in 90~K to 300~K temperature range on warming) was collected at the XPD beamline (28-ID-2) at the NSLS-II at BNL using a similar RAPDF setup but with x-ray wavelength of 0.1901~\AA\ and sample-to-detector distance of 251.1493~mm.
The corresponding instrument resolution parameters were determined to be $Q_{damp} = 0.032$~\AA$^{-1}$ and $Q_{broad} = 0.010$~\AA$^{-1}$, implying similar instrument resolution effects across the two sets of measurements.

The collected x-ray data frames were summed, corrected for detector and polarization effects, and masked to remove outlier pixels before being integrated along arcs of constant momentum transfer $Q$,
to produce 1D powder diffraction patterns using the \pyfai program~\cite{Ashiotisfastazimuthalintegration2015a}. Standardized corrections and normalizations were applied to the data to obtain the reduced total scattering structure function, $F(Q)$, which was Fourier transformed to obtain the PDF, using \pdfgetxthree~\cite{juhas;jac13} within \xpdf~\cite{YangxPDFsuiteendtoendsoftware2015}. The maximum range of data used in the Fourier transform was chosen to be $Q_{max} = 25.0$~\rAA, so as to give the best trade-off between statistical noise and real-space resolution.

{\it Neutron PDF experiment.\textemdash} The time-of-flight (TOF) neutron total scattering measurements were conducted at the NOMAD beamline (BL-1B)~\cite{NeuefeindNanoscaleOrderedMAterials2012} at the Spallation Neutron Source (SNS) at Oak Ridge National Laboratory. The \mto powder sample was loaded into a 3~mm diameter silica capillary mounted on a sample shifter, and data were collected from 100~K to 500~K on warming using an Ar flow cryostream.
The neutron PDF instrument resolution parameters were determined as $Q_{damp} = 0.0232$~\AA$^{-1}$ and $Q_{broad} = 0.0175$~\AA$^{-1}$ by fitting the neutron PDF of a NIST standard 640d crystalline silicon sample.

The neutron data were reduced and transformed to the PDF with $Q_{max} = 25.0$~\rAA\ using the automated data reduction scripts available at the NOMAD beamline.

{\it Structural modeling.\textemdash} The PDF modeling programs \pdfgui and \cmi were used for PDF structure refinements~\cite{farro;jpcm07,juhas;aca15}. In these refinements $U_{iso}$~(\AA$^2$) is the isotropic atomic displacement parameter (ADP) and the ADPs of the same type of atoms are constrained to be the same; $\delta_2$~(\AA$^2$) is a parameter that describes correlated atomic motions~\cite{jeong;jpc99}.
The PDF instrument parameters $Q_{damp}$ and $Q_{broad}$ determined by fitting the PDF from the well crystallized standard sample under the same experimental conditions are fixed in the structural refinements on \mto dataset.

To estimate the lengthscale of local structural correlations, a $r_{min}$ dependent fit is performed on select high temperature x-ray data (300~K, 400~K, and 500~K). During the fit, the average high-temperature cubic \mto model is used, fixing the upper limit of fit range as $r_{max} = 50$~\AA, but changing the lower limit $r_{min}$ from $1 \le r_{min} \le 36$~\AA\ in 0.2~\AA\ $r$-steps. Each fit range uses the same initial parameter values and $\delta_2$ term is not applied since the data fitting range does not consistently include the low-$r$ region in all these refinements.

The Rietveld refinement on the neutron TOF Bragg data was implemented in the GSAS-II software~\cite{TobyGSASIIgenesismodern2013c}.
The instrument parameters were refined to the standard silicon data collected under the same experimental conditions and then fixed in the \mto Rietveld refinements.
The sequential refinement option was used to refine the temperature series dataset collected at $2\theta=120^{\circ}$ detector bank from 100~K to 500~K in a systematic manner.

\begin{figure}[tb]
\begin{centering}
\includegraphics[width=1\columnwidth]{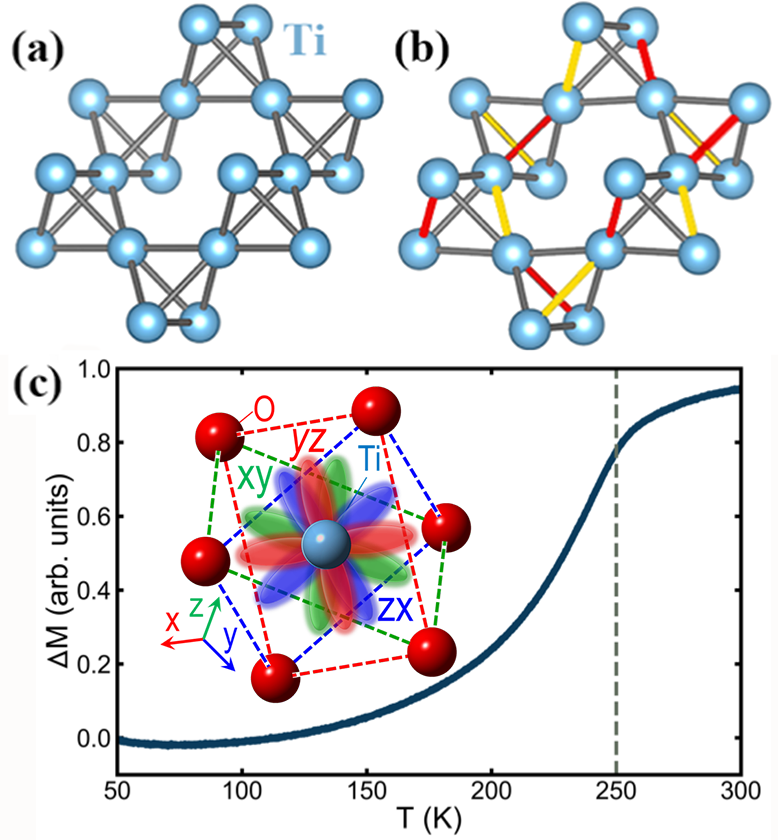}\\
\end{centering}
\caption{
(a,b) The Ti pyrochlore sublattice of corner-shared Ti$_4$ tetrahedra for (a) undistorted (cubic) and (b) distorted (tetragonal) \mto structures, respectively, highlighting both short (red) and long (yellow) Ti-Ti bonds in the distorted structure. (c) The temperature dependence of magnetization with the Curie-Weiss behavior contribution subtracted. The inset shows a TiO$_6$ octahedron and the t$_{2g}$ orbitals that point towards O-O edges. Ti atom is in blue and O atom is in red. The vertical grey dashed line at 250~K marks the MIT temperature.
}
\label{fig;stru_mt_char}
\end{figure}

{\it Structural models.\textemdash} Two candidate \mto models were fit against the experimental data.
In the cubic \mto model (space group: $Fd\overline{3}m$), the atoms sit at the following Wyckoff positions: Mg at 8a (0.125,0.125,0.125), Ti at 16d (0.5, 0.5, 0.5), and O at 32e ($x$, $x$, $x$). The initial lattice parameters and atomic positions are $a=8.509027$~\AA\ and O at (0.25920, 0.25920, 0.25920)~\cite{SchmidtSpinSingletFormation2004}.
The Ti pyrochlore sublattice is shown in Fig.~\ref{fig;stru_mt_char}(a), indicating that all the Ti-Ti bonds are of equal length, reflecting regular Ti$_{4}$ tetrahedra.
In the tetragonal \mto model (space group: $ P4_12_12$), the atoms sit at the following Wyckoff positions: Mg at 4a ($x$,$x$,0), Ti at 8b ($x$, $y$, $z$), O1 at 8b ($x$, $y$, $z$), and O2 at 8b ($x$, $y$, $z$). The initial lattice parameters and atomic positions are $a=6.02201$~\AA, $c=8.48482$~\AA, Mg at (0.7448, 0.7448, 0), Ti at (-0.0089, 0.2499, -0.1332), O1 at (0.4824 0.2468 0.1212), and O2 at (0.2405, 0.0257, 0.8824)~\cite{SchmidtSpinSingletFormation2004}.
The corresponding distorted Ti sublattice is presented in Fig.~\ref{fig;stru_mt_char}(b), showing that one Ti-Ti bond gets shorter (indicated in red) and one gets longer (in yellow) out of the six Ti-Ti bonds of each Ti$_{4}$ tetrahedron.
The lattice parameters of these two models have the relationship of $c_{c} \sim c_{t} \sim \sqrt{2}a_{t}$, where the subscripts $c$ and $t$ refer to the cubic and tetragonal models, respectively.

\section{Vanishing spin singlet dimers}


\subsection{Canonical behavior}

The magnetization, $M(T)$, of the sample up to 300~K is shown in Fig.~\ref{fig;stru_mt_char}(c).
A low temperature Curie-Weiss like component has been subtracted to account for the effect of magnetic impurities contributing to the signal at low temperature. The onset of a broad dimerization transition is observed at around 250~K, which is close to the literature reported MIT temperature of 260~K~\cite{IsobeObservationPhaseTransition2002,VasilievSpecificheatmagnetic2006} and implies canonical behavior of our sample.

\begin{figure}[tb]
\begin{centering}
\includegraphics[width=1\columnwidth]{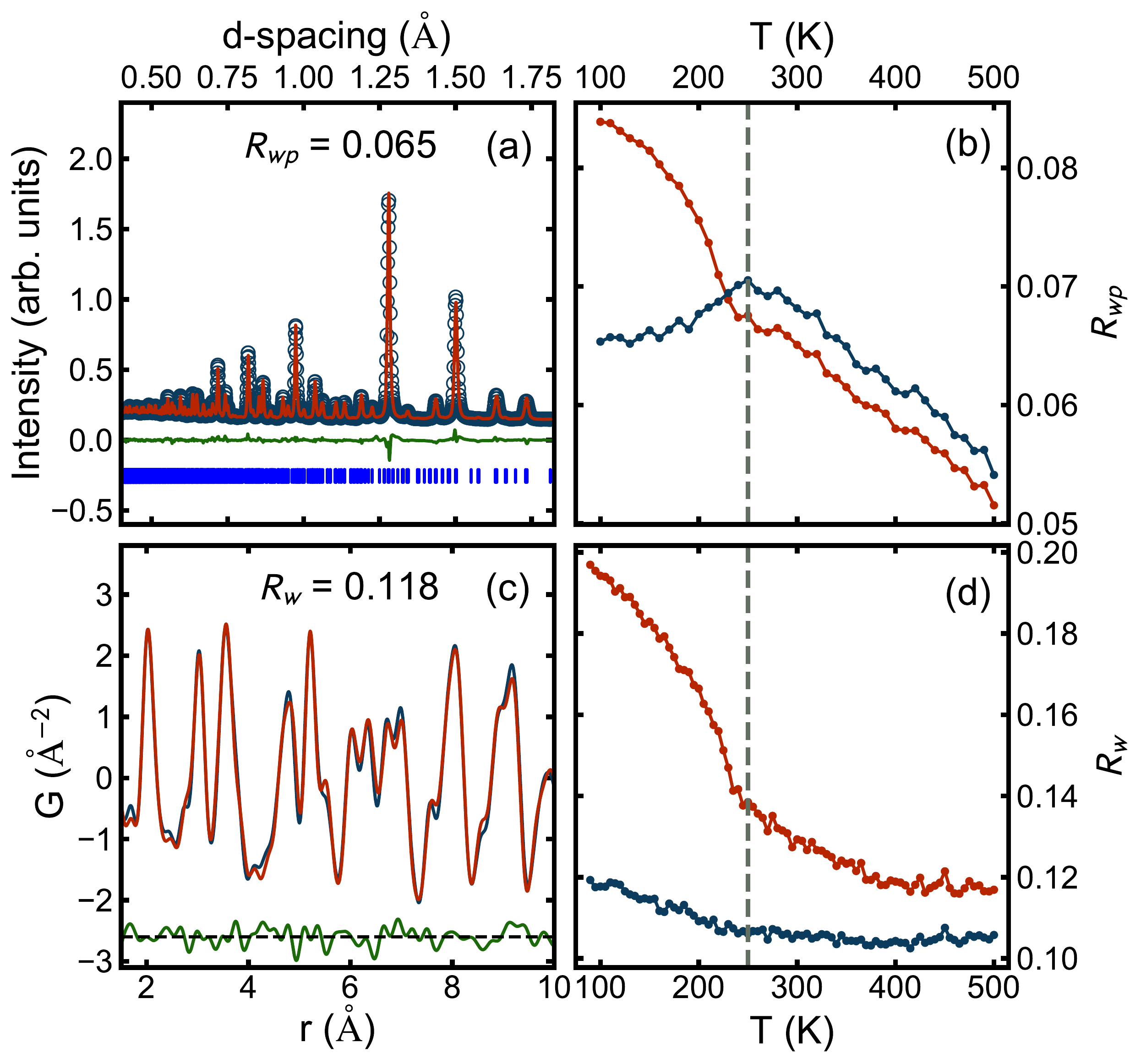}\\
\end{centering}
\caption{
(a) The 100~K neutron powder diffraction pattern (blue) collected using $2\theta=120^{\circ}$ detector bank fit by tetragonal \mto model (red) with difference curve (green) offset below.
(b) The resulting neutron Rietveld refinement weighted profile agreement factor $R_{wp}$ values versus temperature on neutron TOF Bragg data using tetragonal (blue) and cubic (red) \mto models from 100~K to 500~K (120$^{\circ}$ detector bank data).
(c) The x-ray PDF of 100~K data (blue) fit by tetragonal model (red) over the range of $1.5<r<10$~\AA. The difference curve (green) is shown offset below.
(d) The resulting x-ray PDF refinement goodness-of-fit parameter $R_w$ values versus temperature on x-ray data using tetragonal (blue) and cubic (red) \mto models from 100~K to 500~K over the range of $1.5<r<10$~\AA. The vertical grey dashed line at 250~K indicates the MIT temperature.
}
\label{fig;bank4_4panel_fits_rw_versus_pdf}
\end{figure}

We next establish that the average structural behavior of our sample agrees with other observations in the literature~\cite{IsobeObservationPhaseTransition2002,SchmidtSpinSingletFormation2004,TalanovTheorystructuralphase2015} by carrying out Rietveld refinements fitting tetragonal and cubic \mto models to the neutron time-of-flight (TOF) Bragg data measured from 100~K to 500~K.
The refinement results from select fits are reproduced in Table~\ref{tab;fit_rietveld} presented in Appendix~\ref{sec_supplementalResults}, and a representative fit, of the tetragonal model to the 100~K dataset, is shown in Fig.~\ref{fig;bank4_4panel_fits_rw_versus_pdf}(a).
The tetragonal distortion, given by $c/\sqrt{2}a$, is small at all temperatures and the data are not of high enough resolution to directly observe distinct peaks that would indicate any tetragonal distortion.  However, the temperature dependence of the weighted profile agreement factor, $R_{wp}$, of the two models is shown in Fig.~\ref{fig;bank4_4panel_fits_rw_versus_pdf}(b), which clearly implicates the tetragonal model as preferred at low-temperature, but not at high temperature, above around 230-250~K, which is close the literature reported MIT temperature 260~K~\cite{IsobeObservationPhaseTransition2002}.

\subsection{Local structure behavior}

We investigate the behavior of the local structure by performing a PDF structural refinement of the x-ray total scattering PDF data from 90~K to 500~K. Fits were carried out over the data range of $1.5<r<4$~\AA\ and $1.5<r<10$~\AA\ and the results are qualitatively the same.
A representative fit of the tetragonal model to the 100~K dataset is shown in Fig.~\ref{fig;bank4_4panel_fits_rw_versus_pdf}(c).  This time, the temperature dependence of the goodness-of-fit parameter, $R_w$, of the tetragonal and cubic models (Fig.~\ref{fig;bank4_4panel_fits_rw_versus_pdf}(d)) indicates that the tetragonal model fits the local structure better than the cubic model at all temperatures.  In the local structure the tetragonal distortion is always evident.

\begin{figure}[tb]
\begin{centering}
\includegraphics[width=1\columnwidth]{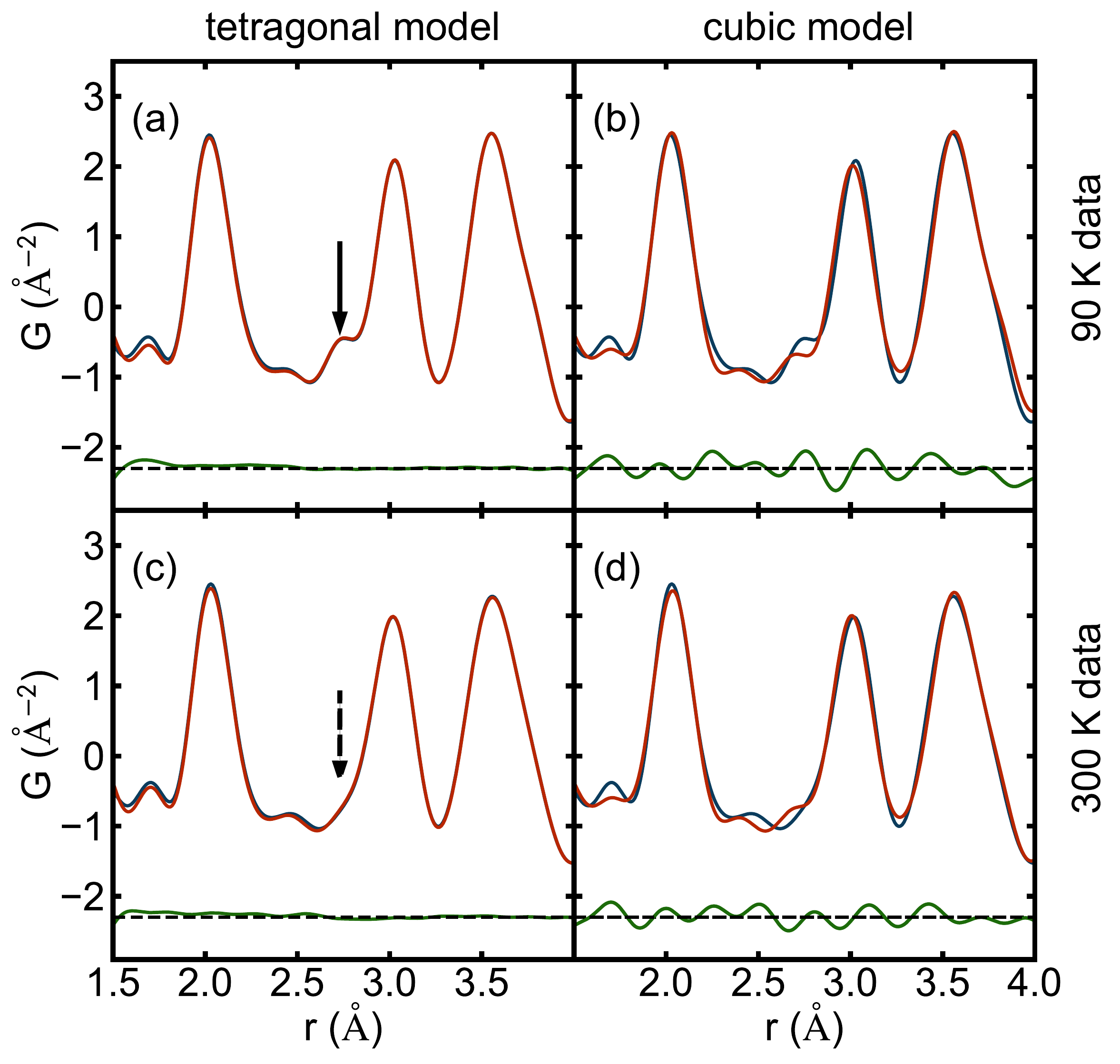}\\
\end{centering}
\caption{(a-d) The x-ray PDF data (blue) collected at the PDF beamline at (a,b) 90~K and (c,d) 300~K fit by (a,c) tetragonal and (b,d) cubic models (red) over the range of $1.5<r<4$~\AA. The difference curves (green) are shown offset below.
}
\label{fig;xray_tetra_cubic_90K_300K_0_4A_fit_may}
\end{figure}

To see visually in the PDF how the cubic model fails we consider the low and high-temperature PDFs and fits with the cubic and tetragonal models, shown in Fig.~\ref{fig;xray_tetra_cubic_90K_300K_0_4A_fit_may}.
The low temperature PDFs (below the MIT) are shown in the top row ((a) and (b)) and the high temperature (above the MIT) in the lower row ((c) and (d)). The tetragonal model fit is shown in the first column ((a) and (c)) and the cubic fit is shown in the second column ((b) and (d)).
At low temperature, as expected, the tetragonal model (Fig.~\ref{fig;xray_tetra_cubic_90K_300K_0_4A_fit_may}(a)) fits much better than the cubic model (Fig.~\ref{fig;xray_tetra_cubic_90K_300K_0_4A_fit_may}(b)).
At 300~K we have already established that the global structure is cubic; however, again we see that in these local structural fits the tetragonal model (Fig.~\ref{fig;xray_tetra_cubic_90K_300K_0_4A_fit_may}(c)) is superior to that of the cubic model (Fig.~\ref{fig;xray_tetra_cubic_90K_300K_0_4A_fit_may}(d)), with smaller oscillations in the difference curve.
To emphasize our argument, we note that the difference curves are similar when data collected at distinct temperatures are fit using identical symmetry constraints. Specifically, careful inspection of the difference curves in Fig.~\ref{fig;xray_tetra_cubic_90K_300K_0_4A_fit_may}(b) and (d) reveals that the positions of residual maxima and minima are nearly identical, although their amplitudes are smaller in Fig.~\ref{fig;xray_tetra_cubic_90K_300K_0_4A_fit_may}(d).
Since this difference signal represents the inability of the cubic model to fit the data, it is further support to the idea that a similar tetragonally distorted structure is present in the low-$r$ region at 90~K and at 300~K, but smaller in amplitude at 300~K.
To validate that the results are reproducible, Fig.~\ref{fig;xray_tetra_cubic_90K_300K_0_4A_fit_may} represents x-ray PDF data collected at the PDF beamline, and data collected from the XPD beamline is shown in Fig.~\ref{fig;xray_tetra_cubic_90K_300K_0_4A_fit} presented in Appendix~\ref{sec_supplementalResults}, which reproduces the same result as discussed above.

In the cubic \mto structure unit cell, all six Ti-Ti bonds have the same length, 3.008~\AA, whereas in the tetragonal \mto model, one of the six Ti-Ti distances is shortened (2.853~\AA), forming dimers, and one Ti-Ti bond is longer (3.157~\AA)~\cite{TalanovTheorystructuralphase2015}.
The dimer contact is considerably shorter than the 2.917~\AA\ found in titanium metal~\cite{Hullcrystalstructurescommon1922}, indicating a strong covalent interaction in \mto.
From the analysis of the average structure, it was understood that the MIT was accompanied by the formation of Ti-Ti structural dimers.  Above we showed the observation from the PDF of local tetragonality at high temperature which may indicate that local dimers survive above the MIT.  This would seem to be in qualitative agreement with a prior observation from a neutron PDF study that reported the persistence of Ti-Ti dimers up to high temperature~\cite{TorigoeNanoscaleicetypestructural2018}.  In this picture the dimers survive locally but become disordered at the transition where the structure becomes globally cubic on warming.
However, our PDF analysis clearly shows that the large amplitude structural dimers actually do disappear at the MIT, as described below, and that the local tetragonality that we observe at high temperature in the PDF has a more subtle origin as we develop in greater detail below.

\subsection{Orbital degeneracy lifting or spin dimerization?}

\begin{figure}[tb]
\begin{centering}
\includegraphics[width=1\columnwidth]{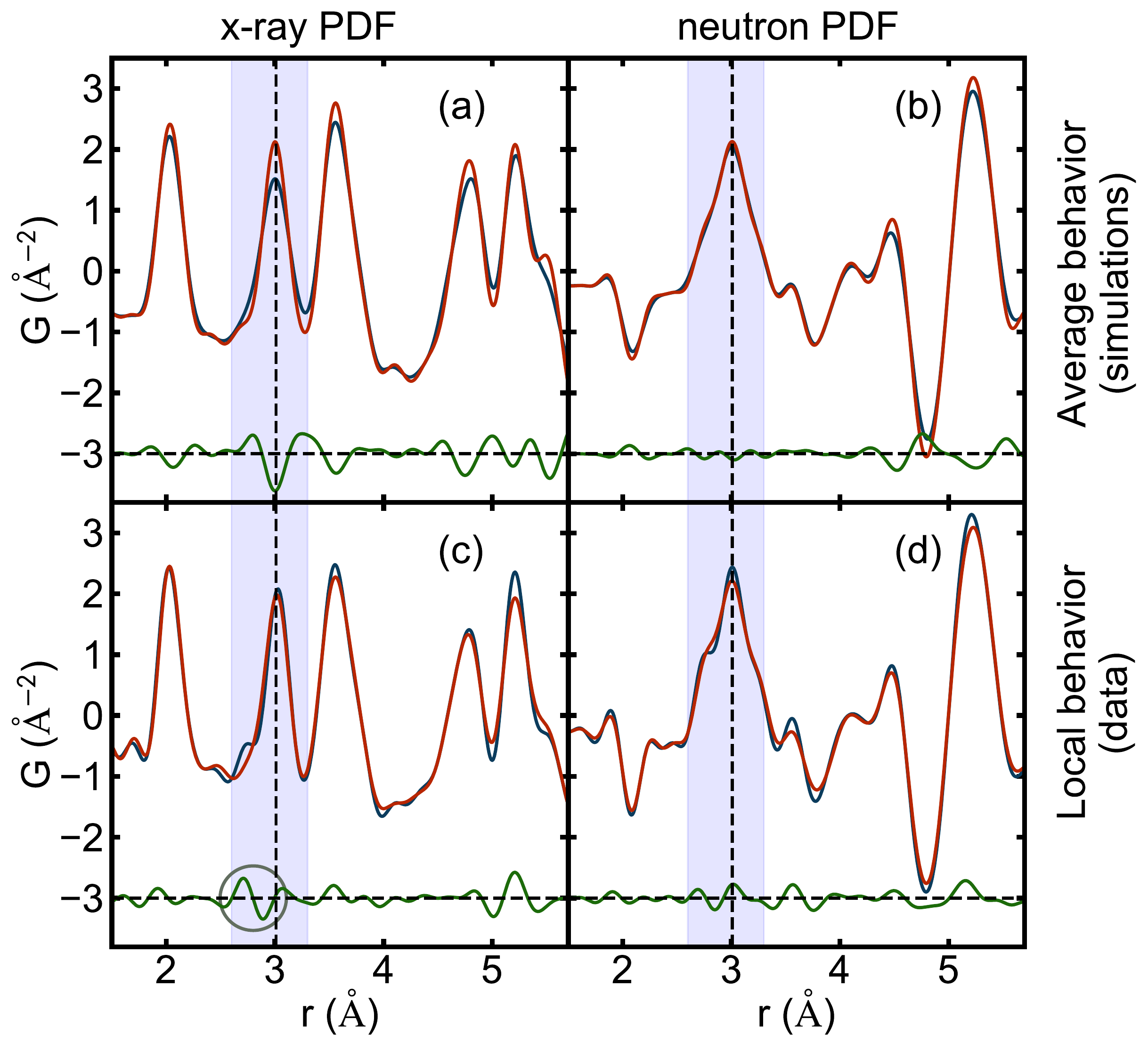}\\
\end{centering}
\caption{(a,b): The simulated (a) x-ray  and (b) neutron PDFs of tetragonal (blue) and cubic (red) \mto models in the low-$r$ region. All atoms use the same isotropic atomic displacement parameter $U_{iso} = 0.005$~\AA$^2$. $Q_{max}$, $Q_{damp}$ and $Q_{broad}$ are set as the same values as the experimental PDFs. The difference curves (tetragonal-cubic) are shown offset below. The nearest Ti-Ti bonds are highlighted in the blue vertical span region.
(c,d): The experimental (c) x-ray and (d) neutron PDFs from \mto collected at 100~K (blue) and 300~K (red). The difference curves (100~K - 300~K) are shown offset below in green.
The vertical dashed lines at $r=3.01$~\AA\ represent the length of the undistorted Ti-Ti bond length in the average cubic model.
}
\label{fig;xray_neutron_100_300K_calc}
\end{figure}

To establish the disappearance of the structural dimers at the phase transition, we simulated x-ray PDFs of the cubic (no dimers) and tetragonal (dimers) models.  These are plotted as the red and blue curves, respectively, in Fig.~\ref{fig;xray_neutron_100_300K_calc} (a).
A number of the PDF peaks are affected on the transition through the MIT, with the largest change observed on the peak at around 3.0~\AA, which contains the shortest Ti-Ti distances.
Based on the change in crystal structure, the expected change in the PDF results in \sout{a }\ins{the} characteristic M-shaped signature in the difference curve \sout{ plotted offset below}\ins{seen in Fig.~\ref{fig;xray_neutron_100_300K_calc} (a)} which comes from the disappearance of the long and short Ti-Ti bonds associated with the dimer, leaving all the Ti-Ti distances the same.
\sout{This can be compared with what happens to the measured PDF data as the sample crosses the MIT, shown in Fig.~\ref{fig;xray_neutron_100_300K_calc}(c) which compares the data-PDFs from \mto collected at 300~K (red) and 100~K (blue).}
\ins{This can be compared to the difference in the measured PDF data as \mto crosses the MIT, shown in Fig.~\ref{fig;xray_neutron_100_300K_calc}(c) at 300~K (red) and 100~K (blue).}
\sout{The first observation is that there is a significant change in the relevant 3.0~\AA\ peak on going through the MIT.}
\ins{First, we see a significant change in the relevant 3.0~\AA\ peak when moving through the MIT.}
This \ins{change} is inconsistent with the \sout{observation}\ins{conclusion} from the previous neutron PDF study that the dimers \ins{are retained,} unaltered\ins{,} to high temperature in the local structure~\cite{TorigoeNanoscaleicetypestructural2018}, \ins{necessitating} an order-disorder scenario in which the local structure across the transition \ins{is unchanged}\sout{would be the same}, with small or no change in the PDF.
Instead we see a significant change in the x-ray PDF at the phase transition, with PDF intensity moving from the short $r$ position \sout{back}towards \ins{that}\sout{the $r$ position} of the average Ti-Ti distance, as indicated by the grey circle in Fig.~\ref{fig;xray_neutron_100_300K_calc}(c).

\begin{figure}[tb]
\begin{centering}
\includegraphics[width=1\columnwidth]{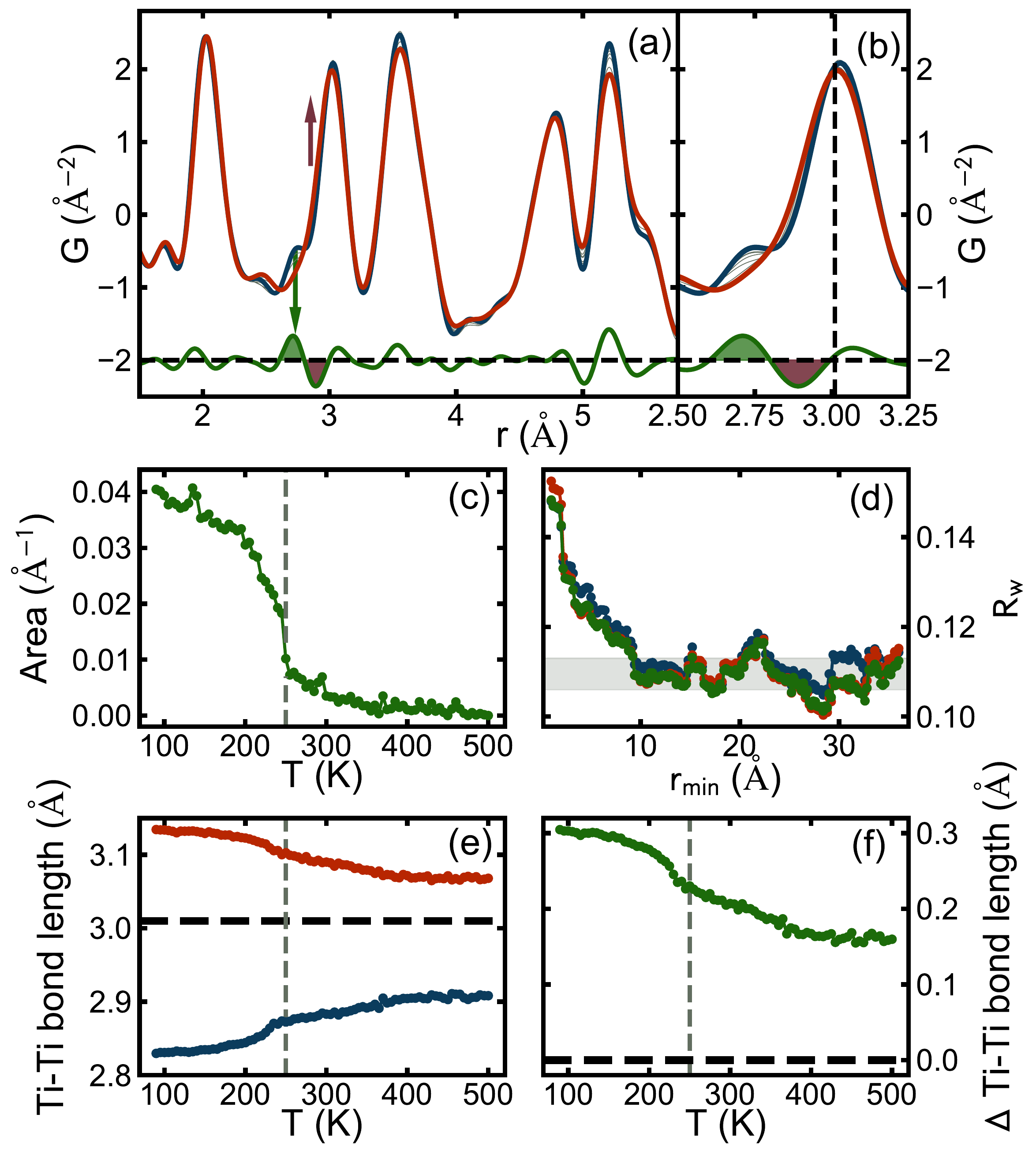}
\end{centering}
\caption{
(a) PDFs measured as a function of temperature from 90~K (blue) to 300~K (red) in the low-$r$ region. The intermediate data-sets are plotted in grey. The vertical green arrow at $r=2.71$~\AA\ and the purple arrow at 2.89~\AA\ represent how the position of the Ti-Ti short bond changes on warming. The difference curve shown offset below is that between the 90~K dataset subtracting the 300~K dataset.  The positive/negative feature between $2.6<r<3.0$~\AA\ indicates that intensity in the PDF at 2.71~\AA\ at low temperature is shifting to the position 2.89~\AA\ at high temperature.
(b) Re-plot of (a) over an expanded $r$-range. The vertical dashed line at $r=3.01$~\AA\ represents the length of the undistorted Ti-Ti bond length in the cubic average structure.
(c) The temperature dependence of the integrated area in the difference curve shown shaded green. The vertical grey dashed line at 250~K indicates the MIT temperature.
(d) The x-ray PDF refinement goodness-of-fit $R_w$ values versus a variable $r_{min}$ (for an $r_{max}$ fixed to 50~\AA) when fitting a cubic \mto model to 300~K (blue), 400~K (green), and 500~K (red) data.  In each refinement $r_{min}$ was allowed to vary from $1 \le r_{min} \le 36$~\AA\ in 0.2~\AA\ $r$-steps. The fits with the high-$r$ yield the behavior of the average structure. The fits with the low $r_{min}$ are weighted by the local structural signal.
(e, f) The temperature dependence of the short (blue) and long (red) Ti-Ti bond lengths, and their differences (green) from PDF structural modeling using the tetragonal structure over the range of $1.5<r<10$~\AA. The horizontal dashed line represents the length of the undistorted Ti-Ti bond length in the cubic average structure. The vertical grey dashed line at 250~K indicates the MIT temperature.
}
\label{fig;xray_90_300K_temp_dimer_1_5A_4panel}
\end{figure}

\sout{Careful inspections show that the difference curve from the experimental PDFs (Fig.~\ref{fig;xray_neutron_100_300K_calc}(c)) is significantly different from that of the calculated PDFs (Fig.~\ref{fig;xray_neutron_100_300K_calc}(a)).}
\ins{It is clear from careful inspection that the observed and simulated difference curves across the transition (Fig.~\ref{fig;xray_neutron_100_300K_calc}(c) and (a), respectively) are significantly different.}
PDF peak intensity redistribution associated with the dimer removal is seen \ins{in the differential curve} as a transfer of intensity \sout{in the differential curve} from 2.71~\AA\ \sout{position moving back to a position that is centered around}\ins{to} 2.89~\AA, i.e., on the low-$r$ side of the average Ti-Ti distance at 3.01~\AA\ marked by the vertical dashed line in the figure.
This implies that the very short Ti-Ti ion dimers disappear at the phase transition, but the \ins{associated PDF} intensity \sout{in the PDF}is shifted \sout{back}to a position that is still shorter than that of the average Ti-Ti distance: shortened Ti-Ti contacts in fact do exist at high temperature above the MIT but they are not the original spin singlet dimers.
Our results are consistent with the \ins{disappearance of} dimers \sout{disappearing}at the phase transition on warming, but \ins{a persistence of a local tetragonality, smaller in magnitude than that associated with the the dimer phase.}\sout{local tetragonality getting smaller than that corresponding to the dimer phase still persisting well above this transition.}
This behavior is shown on an expanded scale in Fig.~\ref{fig;xray_90_300K_temp_dimer_1_5A_4panel}(b) where PDFs measured at multiple temperatures are compared, focusing on the 3~\AA\ peak.
The feature in the difference curve that shows the shift of the intensity associated with the Ti-Ti short bonds from 2.71~\AA\ to the longer position (2.89~\AA) is shown below shaded green (for the loss of short bonds) and purple (for the gain in longer bonds).
This result is qualitatively supported by the PDF structural modeling over the range of $1.5<r<10$~\AA, \ins{shown in Fig.~\ref{fig;xray_90_300K_temp_dimer_1_5A_4panel}(e) and (f),} where the short Ti-Ti bond shifts from 2.83~\AA\ to 2.88~\AA\ over the same temperature range, but never \ins{converges}\sout{returns} to the average cubic value of 3.01~\AA\ at 300~K.
\sout{This is shown in Fig.~\ref{fig;xray_90_300K_temp_dimer_1_5A_4panel}(e) and (f).}

The full temperature dependence of the dimer disappearance may then be extracted by integrating the shaded regions in the difference curve and is shown in Fig.~\ref{fig;xray_90_300K_temp_dimer_1_5A_4panel}(c), where we plot the integrated area shown shaded in green in the difference curve vs. temperature.  In this case the difference is always taken with respect to the 500~K dataset. The intensity decreases gradually until around 200~K, where it rapidly falls off.  The rate of fall-off in this intensity then slows again above 250~K. There are two principal contributions to the differential signal below the transition: the changes in the local structure and the thermal broadening effects. Thermal broadening effects in the differential are gradual and typically small, particularly from one data point to the next corresponding to the trends observed below $\sim$ 200~K and above $\sim$ 250~K. On the other hand, the signal in the differential coming from the dimers is considerably larger, roughly proportional to the big step seen at the transition, and dissipates rapidly as the temperature passes through the MIT range. This is a model-independent way of observing how the local dimer disappears on warming.

\subsection{X-ray PDF versus neutron PDF}

We may seek an explanation for why we see the dimers disappear at the transition from our x-ray PDF measurements, whereas this was not evident in the earlier neutron PDF study~\cite{TorigoeNanoscaleicetypestructural2018}. Observing the local Ti dimer is complicated in the neutron case by the relatively weak and negative neutron scattering length of titanium, and an appreciable overlap of titanium contributions with strong oxygen contributions. To show this, we perform the same comparison that we just made for the x-ray PDFs, but on neutron data.  The neutron simulations are shown in Fig.~\ref{fig;xray_neutron_100_300K_calc}(b) and  the neutron experimental PDFs in Fig.~\ref{fig;xray_neutron_100_300K_calc}(d).  It is clear that the signals in the difference curve, both for the simulations based on the average structure and the data PDFs themselves, are much smaller than for the x-ray case. This is because the most important signal in \mto is coming from the Ti ions that are relatively strong scatterers in the x-ray case but not in the neutron measurement. Specifically, in the x-ray case, Ti scatters 2.75 times stronger than O and 1.8 times stronger than Mg, whereas for neutrons the scattering of Ti is 1.7 times weaker than that of O and 1.6 times weaker than that of Mg. In addition, the neutron PDF intensity in the range of interest is dominated by the contributions from O-O pairs constituting TiO$_{6}$ octahedra, evidenced in the data as the additional shoulder intensity features around the 3~\AA\ peak (Fig.~\ref{fig;xray_neutron_100_300K_calc}(d)), as compared to the x-ray case where such features are largely absent (Fig.~\ref{fig;xray_neutron_100_300K_calc}(c)). This may explain the different interpretation in the earlier neutron PDF study~\cite{TorigoeNanoscaleicetypestructural2018}.  However, the x-ray data unambiguously show the disappearance of the full-amplitude spin singlet Ti-Ti dimers.

\subsection{Detection of the ODL state}

This behavior, where sizeable distortions associated with ordered spin singlet dimers evolve into smaller local distortions with spin singlets disassembled, is reminiscent of that observed in the \cis system.  In that case a local symmetry broken orbital-degeneracy-lifted (ODL) state was observed up to the highest temperatures studied~\cite{BozinLocalorbitaldegeneracy2019c}.  On cooling these local distortions of broken symmetry ordered into a long-range orbitally ordered (LROO) state.  Only below the LROO transition did charges disproportionate, forming a charge density wave accompanied by a Peierls distortion and the formation of Ir-Ir dimers with very short Ir-Ir bonds.  We believe our observations in \mto suggest a similar kind of ODL behavior, which we explore to a greater extent below, though different in detail because of the different charge filling.

As discussed above, we can rule out that the local structure is changing in the same way as the average structure.  The {\it expected} changes in the PDF at low-$r$ due to average structure changes at the MIT make all the low-$r$ peaks sharper for $T>T_{MIT}$, as shown in Fig.~\ref{fig;xray_neutron_100_300K_calc}(a). However, the data do not show this (Fig.~\ref{fig;xray_neutron_100_300K_calc}(c)). There is a small change in the local structure, evidenced by the feature in the difference curve around 2.9~\AA, indicated by the grey circle in Fig.~\ref{fig;xray_neutron_100_300K_calc}(c), but it is smaller than the average structure change at the MIT.
As discussed above, the short Ti-Ti dimers ($r=2.71$~\AA\ shoulder) go away on warming by a shift to a longer bond (around $r=2.89$~\AA), but this ``longer" bond is still shorter than the average 3.01~\AA\ Ti-Ti bond distance expected from the cubic average model.
These two behaviors exactly mimic the ODL state found in \cis~\cite{BozinLocalorbitaldegeneracy2019c}.

As is evident in Fig.~\ref{fig;bank4_4panel_fits_rw_versus_pdf}(d), the tetragonal model fits the local structure better than the cubic model at all temperatures to 500~K, the highest measurement temperature. As in \cis~\cite{BozinLocalorbitaldegeneracy2019c} the dimers disappear at the MIT transition on warming, but the local symmetry broken state with, presumably, fluctuating short Ti-Ti bonds is present to high temperature.

\subsection{Spatial extent of the ODL state}

It is of interest to explore whether the fluctuating short Ti-Ti bonds at temperature above the MIT correlate with each other, and how this varies with temperature.  To extract the correlation length of the local fluctuating \sout{ODL}\ins{symmetry broken} states, $r_{min}$ dependent PDF fits were performed on selected high-temperature datasets (300~K, 400~K, and 500~K).  In these models the high-temperature average cubic model was fit over a range from $r_{min}$ to a fixed $r_{max}=50$~\AA, where $r_{min}$ was allowed to vary from $1 \le r_{min} \le 36$~\AA . When $r_{min}$ is large the fit is over just the high-$r$ region of the PDF and will retrieve the average structure and the cubic fit will be good.  As $r_{min}$ extends to lower values, progressively more of the local structure that has a tetragonal distortion is included in the fit and the agreement of the cubic model becomes degraded.  The resulting $R_w(r_{min})$ is shown in  Fig.~\ref{fig;xray_90_300K_temp_dimer_1_5A_4panel}(d).
The cubic fits are good over the entire region, $r_{min}>10$~\AA, with very little variation in $R_w$ indicated by the horizontal grey band in the figure, but rapidly degrade below this length-scale.
This suggests that the \sout{ODL}\ins{symmetry broken local} distortions have a correlation length of around 1~nm but that this correlation length does not vary significantly in the temperature range above 300~K. This agrees well with the previously reported length-scale of local tetragonality based on neutron PDF analysis~\cite{TorigoeNanoscaleicetypestructural2018}.

%

\section{Two-orbital ODL state}

\subsection{The ODL regime}

%
The driving force behind the high temperature local symmetry breaking in \cis was shown to be of electronic origin involving orbital degeneracy lifting~\cite{BozinLocalorbitaldegeneracy2019c}. Significantly, the isostructural and isoelectronic sister compound, \cise, remains metallic down to the lowest temperature and does not show local symmetry breaking orbital-degeneracy-lifted (ODL) effects. This intriguingly suggests that the ODL state could be {\it a prerequisite} for the MIT~\cite{BozinLocalorbitaldegeneracy2019c} in these spinels. Here we argue that also in \mto local ODL effects are present and produce the local symmetry breaking in the high temperature region that we report here.  By analogy with \cis, the ODL states are precursor states to the spin singlet dimerization and MIT.

\begin{figure}[tb]
\begin{centering}
\includegraphics[width=1.0\columnwidth]{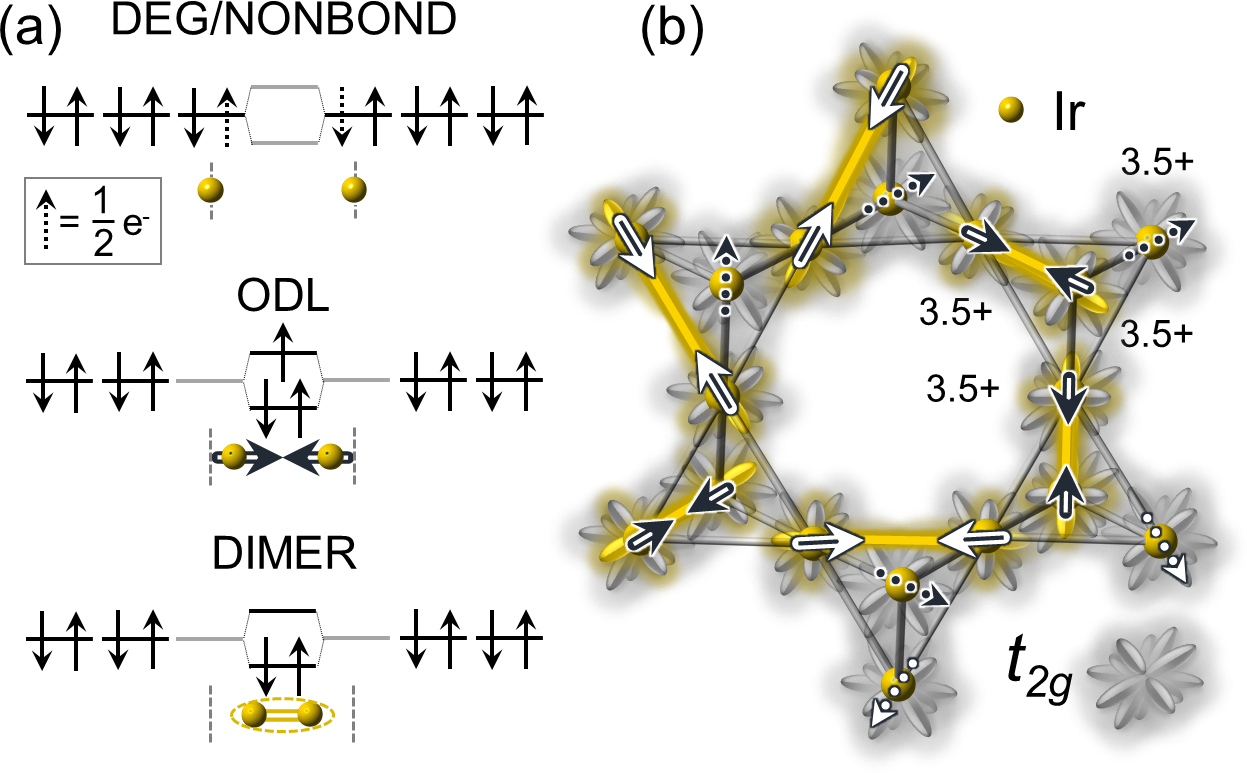}
\end{centering}
\caption{Recapitulation of the ODL mechanism in \cis thiospinel: (a) The energy diagram of atomic and molecular orbitals for nearest Ir atom pairs in various situations: degenerate/nonbonding (DEG/NONBOND, top), orbital degeneracy lifted (ODL, middle), and spin singlet dimer (DIMER, bottom).
Horizontal arrows indicate the ODL displacements.
(b) The ODL state in \cis in \ttg orbital manifold representation on a segment of pyrochlore sublattice.
The ODL active \ttg orbitals are shown in yellow, passive orbitals are shown in grey.
Each Ir atom participates in exactly one ODL state. The ODL states are randomly placed following principles discussed in text.
The arrows denote the displacements of Ir associated with the ODL states and provide mapping onto the ``two in -- two out" ice rules.
The white arrows represent in-plane displacements, whereas the blue arrows represent displacements with an out-of-plane component.
The dotted arrows mark the displacements associated with the ODL states occurring in neighboring tetrahedra that are not represented in the displayed structural segment.
}
\label{fig;cis_odl_review}
\end{figure}
%

\begin{figure*}[tb]
\begin{centering}
\includegraphics[width=1.8\columnwidth]{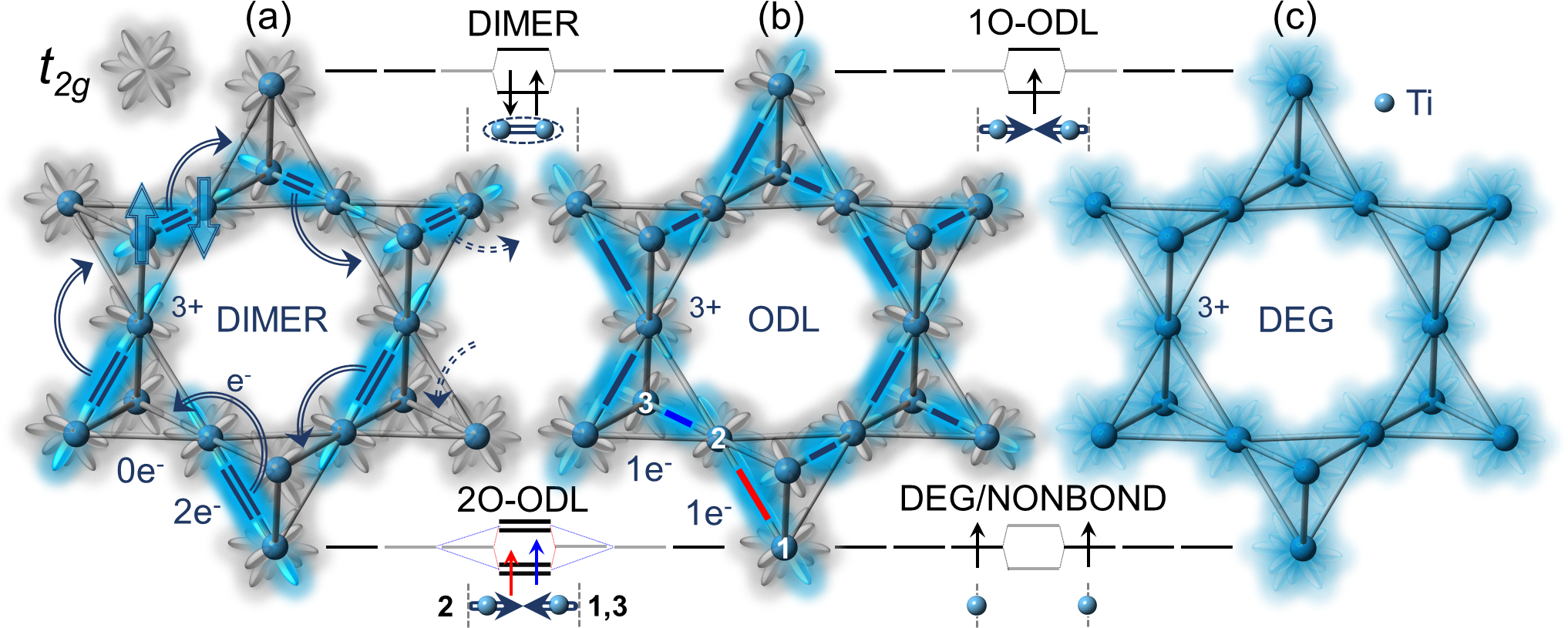}
\end{centering}
\caption{The Ti \ttg orbital manifold on a pyrochlore sublattice of \mto spinel: (a) Spin-singlet dimerized lattice within the tetragonal model; (b) ODL lattice within the tetragonal model; (c) Lattice of degenerate \ttg orbitals as portrayed by the cubic model.
The charge state of all Ti in all models is +3.
Electron density in \ttg orbitals is indicated by color (grey is empty) and its distribution is proportional to the blue color intensity.
The Ti dimers in (a) are indicated by the double blue lines, while a pair of antiferromagnetically coupled block arrows on an exemplar dimer denotes its spin singlet character.
Note that all Ti are involved in dimerization but only dimers contained in a section of one structural slab are indicated in (a).
Out of all Ti-Ti contacts dimerized contacts have nominal bond charge of 2$e^{-}$, whereas the other contacts carry no net charge in this picture.
The ODL states in (b) are indicated by the single blue lines.
Arc arrows in (a) denote bond charge transfer as the spin singlet dimers convert to the ODL states, according to the model discussed in text.
The insets between the panels provide the energy diagram of atomic and molecular orbitals for nearest Ti atom pairs in various situations: degenerate/nonbonding (DEG/NONBOND, bottom right), spin singlet dimer (DIMER, top left), one-orbital ODL (1O-ODL, top right) and two-orbital ODL (2O-ODL, bottom left).
Note: the 2O-ODL state refers to two orbitals of the same Ti (e.g. one labelled ``2") that are engaged in two 1O-ODL states with two {\it different} Ti neighbors (e.g. these labelled ``1" and ``3"). Differently colored spins and lines denoting ODL contacts indicate this two-component aspect in the schematics for 2O-ODL. The local spin arrangement shown is for illustration purpose only and does not represent an experimentally established alignment.
}
\label{fig;mto_mechanism}
\end{figure*}

The long range pattern of ordered spin singlet dimers differs in detail in the two systems, being octamers in \cis~\cite{RadaelliFormationisomorphicIr32002b} and helices in \mto~\cite{SchmidtSpinSingletFormation2004}, and is presumably dictated by delicate energetics~\cite{DiMatteoValencebondcrystallattice2005,RahamanTetramerorbitalordering2019}. Although intriguing, these intricate aspects of long range order are not of concern here. Our interest lies with internal dimer structure, the relationship between the dimer state and the ODL state, and the electron-hole symmetry that links the two systems~\cite{KhomskiiOrbitallyInducedPeierls2005a}.

Prior to pursuing the analogy between the \cis\ and \mto, we appraise the two systems from the perspective of a regime under which the ODL state can form in a transition metal system~\cite{BozinLocalorbitaldegeneracy2019c}.
Three criteria need to be fulfilled:
(i) {\it filling} -- the transition metal in the system possesses partially filled $d$ orbitals,
(ii) {\it symmetry} -- the high crystallographic point symmetry of the system imposes the orbital degeneracy, and
(iii) {\it topology} -- the structural topology promotes adequate highly directional orbital overlaps.
In practice, materialization of the ODL state could further be impacted by the existence of other competing degeneracy lifting channels that may be available to the system, such as the relativistic spin orbit coupling, the effects of crystal field, etc.
Eventually, formation of the ODL local broken symmetry state is accompanied by associated local structure distortion.
%
Both \cis\ and \mto\ systems meet the criteria: partial filling (Ir is $5d^{5.5}$, Ti is $3d^{1}$), degeneracy promoting high symmetry structure (cubic spinel structure imposes 3-fold degeneracy of the \ttg manifolds in both materials), and favorable structural topology (edge shared IrS$_{6}$ and TiO$_{6}$ octahedra fostering direct \ttg overlap). In both systems short nearest neighbor transition metal contacts are observed in the high temperature metallic regime, that are distinct from the very short spin singlet dimer bonds observed in the insulating state at low temperature.

In what follows we first review the ODL state and the dimerization mechanism within the local picture of \cis, summarized in Fig.~\ref{fig;cis_odl_review}. We then utilize the analogies between the two systems to put forward a scenario portraying the high temperature state and the dimerization mechanism in \mto, illustrated in Fig.~\ref{fig;mto_mechanism}.

\subsection{Single-orbital ODL state}

In \cis the dimers involve two Ir$^{4+}$ ions in $5d^{5}$ configuration with hole character~\cite{BozinLocalorbitaldegeneracy2019c}.
In the localized electron picture, the ODL state is based on molecular orbital (MO) concepts. In this, two neighboring transition metal ions with partially filled degenerate orbitals form a MO state with shared electrons (holes) that lifts the degeneracy and lowers the system energy. In the high temperature regime of \cis iridium nominally has a half-integer valence (3.5+)~\cite{MatsunoPhotoemissionstudymetalinsulator1997b,YagasakiHoppingConductivityCuIr2S42006b}. This means that each Ir$^{3.5+}$ on the pyrochlore sublattice is in a nominal $5d^{5.5}$ state corresponding to half a hole per three degenerate \ttg orbitals (Fig.~\ref{fig;cis_odl_review}(a), top). The sublattice geometry fosters direct overlaps of orbitals from the \ttg manifolds (Fig.~\ref{fig;cis_odl_review}(b)), which, in turn allows neighboring Ir pairs to form a bound state sharing a single hole in the antibonding MO~\cite{BozinLocalorbitaldegeneracy2019c} (middle panel of Fig.~\ref{fig;cis_odl_review}(a)).

Importantly, for any choice of two neighboring Ir on a pyrochlore lattice only one member of the three \ttg orbitals overlap (e.g. {\em xy} with {\em xy}, etc.) along the Ir$_{4}$ tetrahedral edges. Due to the specifics of filling (0.5 holes/Ir) each Ir participates in exactly one such paired state at a time. We call this a {\it one-orbital ODL (1O-ODL)} state. The ODL state is hence comprised of two atomic orbitals, one from each Ir in the pair, with on average 1.5 electrons (0.5 holes) per Ir, resulting in MO with 3 electrons and one hole, as shown in Fig.~\ref{fig;cis_odl_review}(a), with a net spin of 1/2. This configuration results in the observed contraction of the Ir-Ir separation in the local structure vis-{\`a}-vis that expected if orbital degeneracy is retained. Since each iridium has 6 Ir neighbors to pair with, the ODL state fluctuates spatially and, presumably, temporally among ({\em xy, xy}), ({\em yz, yz}), and ({\em zx, zx}) variants, which results in an undistorted cubic structure on average~\cite{BozinLocalorbitaldegeneracy2019c}. One such configuration is illustrated in Fig.~\ref{fig;cis_odl_review}(b), with strong short-range correlations governed by the ``one ODL state per Ir" and the Coulomb bond charge repulsion principles, resulting in a single ODL state per Ir$_{4}$ tetrahedron. Pursuant to this, the ODL state in \cis follows the ``two in -- two out" ice rules~\cite{ander;pr56}, which will be addressed later.

In this localized picture, the ODL state is also a precursor for the spin singlet dimer. The dimer state is attained by removing an excess electron from the antibonding MO of the ODL state, thus stabilizing the bond, as shown in the bottom panel of Fig.~\ref{fig;cis_odl_review}(a). The process involves charge transfer between two ODL Ir$^{3.5+}$-Ir$^{3.5+}$ pairs, one of which becomes dimerized Ir$^{4+}$-Ir$^{4+}$ by losing an electron (or gaining a hole, hence hole dimer) and the other becomes non-ODL (and non-dimer) Ir$^{3+}$-Ir$^{3+}$ by gaining an electron ($d^{6}$-$d^{6}$ configuration).

\subsection{Two-orbital ODL state}

We now turn to \mto\ starting from the local view of spin singlet dimers. Dimerization in \mto, depicted using the Ti \ttg manifold representation, is shown in Fig.~\ref{fig;mto_mechanism}(a), overlaying a fragment of pyrochlore sublattice as seen in the $P4_12_12$ model. In \mto the dimers involve two Ti$^{3+}$ in $3d^{1}$ configuration with a single electron in \ttg manifold, hence Ti$^{3+}$-Ti$^{3+}$ dimers inevitably have electron character by construction. These result in short Ti-Ti dimerized contacts observed in the tetragonal structure at low temperature. Each dimer carries 2$e^{-}$ of net charge. Since each Ti participates in a dimer, and since there is exactly one dimer per Ti$_{4}$ tetrahedron~\cite{DiMatteoValencebondcrystallattice2005,TorigoeNanoscaleicetypestructural2018}, nominal charge count results in 1$e^{-}$/Ti site. This is consistent with no CO being observed experimentally in \mto, in contrast to \cis, implying that all Ti sites are equivalent in this regard~\cite{SchmidtSpinSingletFormation2004}. From the average structure perspective and in the localized picture, supported by the experimentally observed paramagnetism and poor metallic conduction, above the MIT the dimers could be seen to disassemble in such a way as to statistically distribute 1$e^{-}$ of charge evenly across the three degenerate \ttg orbitals, resulting in a cubic structure with all Ti-Ti nearest neighbor contacts equivalent, as schematically presented in Fig.~\ref{fig;mto_mechanism}(c). This implies that, despite nominal charge equivalence of all Ti sites and no site CO observed, some charge transfer, presumably involving bond charge, still has to take place at the transition. This then inevitably implies that the ground state has to involve the bond charge order which coincides with, and is hence indistinguishable from, the observed dimer order in the diffraction measurements.

The implication of charge being equally distributed across the triply degenerate \ttg manifold of Ti in a manner depicted in Fig.~\ref{fig;mto_mechanism}(c), presented also in associated energy diagram (bottom right inset to the Figure), would be that the pyrochlore sublattice is comprised of regular Ti$_{4}$ tetrahedra with equidistant Ti-Ti contacts corresponding to equal bond charge, as described within the cubic spinel model. However, this is not what is observed experimentally. The observations based on the PDF analyses clearly demonstrate that the pyrochlore sublattice is comprised of locally distorted Ti$_{4}$ tetrahedra with a distribution of distances, suggesting that the bond charge remains inequivalent above the MIT. One possibility is that the dimers indeed persist in the metallic regime, as hinted at in the earlier neutron study~\cite{TorigoeNanoscaleicetypestructural2018}.
However, such an interpretation would be inconsistent with magnetization measurements of \mto that establish the disappearance of spin singlets above the MIT.
It would further be inconsistent with the elongation of the short Ti-Ti dimer contacts evidenced in our x-ray PDF analysis, which implies disappearance of spin singlet dimers at MIT even locally. In analogy with \cis, it is plausible that the dimer state gets replaced locally by an ODL-type state in the high temperature metallic phase. Observation of heterogeneous local Ti-Ti contacts corresponding to inequivalent bond charge is consistent with an ODL-like state in \mto. Since the local structures at high temperature and in the ground state are distinct, the MIT in \mto cannot be assigned to a trivial order-disorder type.

Importantly, as careful assessment shows, the ODL state in \mto cannot be exactly mapped onto the ODL state seen in \cis, since the charge filling is different in the two systems (1 $e^{-}$/Ti in \mto, 0.5 holes/Ir in \cis). In the latter case two Ir neighbors can reduce the energy of the system by forming a 1O-ODL state accommodating a common hole, as described in the previous section. This results in a 3-electron state shown in the middle panel of Fig.~\ref{fig;cis_odl_review}(a). Given that the destruction of the spin singlet dimer in \mto involves removal of one of the electrons from the dimer state sketched in the energy diagram in the top left inset to Fig.~\ref{fig;mto_mechanism}, consistent with destabilization and elongation of the short Ti-Ti bond, in the ODL picture the one electron left behind would indeed result in an electron-hole antipode of the 1O-ODL state seen in \cis, as shown in the top right inset to Fig.~\ref{fig;mto_mechanism}. The issue arises with the placement of the extra electron from the dimer. In case of \cis, and due to filling, the removal of one of the two holes constituting a dimer results in generation of two 1O-ODL states on two different pairs of Ir. There, since only 50\% of Ir participates in dimerization, in terms of filling one should think of this process as a replacement of --4+--4+--3+--3+-- charge tetramer with a pair of --3.5+--3.5+-- ODL states achieved by hole redistribution. Since there is no site charge disproportionation in \mto and since each Ti is involved in dimerization, the dimer density per formula unit is twice as large in \mto as in \cis and each Ti has to participate in two independent 1O-ODL states simultaneously. Notably, the geometry of \ttg orbital overlaps imposes a constraint that the two 1O-ODL states for each Ti have to be assembled with two {\it different} Ti neighbors, as illustrated in Fig.~\ref{fig;mto_mechanism}(a),(b). There, for example, the dimer in Fig.~\ref{fig;mto_mechanism}(a) involving Ti labeled ``2" in Fig.~\ref{fig;mto_mechanism}(b) disassembles at MIT to make two 1O-ODL states, one with the Ti neighbor labeled ``1" and another with the neighbor labeled ``3". This results in two short Ti-Ti contacts that are both longer than the dimer distance, but shorter than the average Ti-Ti separation in the cubic structure. We call this a two-orbital ODL state (2O-ODL) given that two atomic orbitals of a single Ti ion are utilized. Such 2O-ODL state is hence comprised of a superposition of two 1O-ODL states of different variety (e.g. ({\em xy, xy}) and ({\em yz, yz}), etc.) that point in different directions and lie along different edges of the pyrochlore sublattice. These ODL distances are marked as thick lines color coded as red and blue in Fig.~\ref{fig;mto_mechanism}(b), and the corresponding state is schematically shown on the energy diagram in the bottom left inset to the Figure, where the matching color coding of the electron spins signifies that they belong to different 1O-ODL states.

The existence of 2O-ODL state in \mto is corroborated by another experimentally observed difference between the two systems: Spatial extent of the local order associated with the metallic regime of \mto is observably larger than that seen in \cis~\cite{BozinLocalorbitaldegeneracy2019c}. The extended character of the local structural distortions associated with the 2O-ODL state in \mto are expected for the following reasons. First, due to filling the two ingredient 1O-ODL prong states in the 2O-ODL super-state cannot both be of the same type (e.g. if one is ({\em xy, xy}), the other can only be ({\em yz, yz}) or ({\em zx, zx})) which would tend to increase local structural correlations. Second, the bond charge on different 1O-ODL bonds is of the same sign, resulting in the Coulomb repulsion which would also maximize the span of the 2O-ODL state itself. The observed local distortions over the length scale of $\sim$1~nm, spanning approximately three Ti$_{4}$ tetrahedra ~\cite{TorigoeNanoscaleicetypestructural2018}, are therefore consistent with the presence of the 2O-ODL state in \mto.

The energetic benefit of the 2O-ODL state over the local spin singlet dimer state is not apparent and remains elusive. The local 2O-ODL and disordered dimer states would both increase the system entropy and, in turn, the entropic contribution to energy would stabilize corresponding short range ordered state at elevated temperature. On the other hand, some vestigial magnetic correlations could be expected in the high temperature ODL regime. Another possibility then is that the principal stabilization of the 2O-ODL state stems from its presumed magnetism. The character of local spin correlations within the 2O-ODL state cannot be established by the analysis carried out in this work. In the 2O-ODL diagram shown in Fig.~\ref{fig;mto_mechanism} the two spins are arbitrarily drawn as parallel to avoid any confusion with the dimer state, but their relationship in fact has not been established experimentally.
In fact, the magnetic response of \mto above the MIT, Fig.~\ref{fig;stru_mt_char}(c), is neither Pauli-like nor Curie-Weiss-like. Rather it resembles that of charge density wave systems at high temperature~\cite{BenchimolAnisotropymagneticsusceptibility1978}, the regime associated with a pseudogap in the electronic density of
states~\cite{BorisenkoPseudogapChargeDensity2008b}.
In these systems magnetic susceptibility behavior at T$>$T$_{s}$ was attributed to fluctuations of the charge density wave amplitude~\cite{JohnstonThermodynamicsChargeDensityWaves1984,ZhangIntertwineddensitywaves2020}. It is thus tempting to speculate that in \mto\ the observed magnetic response may similarly be due to ODL fluctuations.
It would therefore be of appreciable interest to explore this aspect of the 2O-ODL state by techniques sensitive to local magnetism, such as $\mu$SR~\cite{McKenziepositivemuonmSR2013} and magnetic PDF~\cite{FrandsenMagneticpairdistribution2014}, which would be particularly informative in that regard.

While the PDF probe used here provides the information on instantaneous atomic structure, and as such does not differentiate between static and dynamic disorder, the ODL state in these systems is expected to be dynamic. Spatiotemporal fluctuations then average out to perceived undistorted cubic average structure as observed crystallographically.
Notably, the resistivity just above the MIT in \cis\ is about 2~m$\Omega$cm and linearly increasing with temperature~\cite{BurkovAnomalousresistivitythermopower2000b}, ascribed to bipolaronic hopping mechanism~\cite{YagasakiHoppingConductivityCuIr2S42006b,TakuboIngapstateeffect2008b}, whereas in the metallic regime of \mto\ electric resistivity is not only substantially higher, but it decreases with increasing temperature in an insulator-like manner~\cite{IsobeObservationPhaseTransition2002}. This stark difference in the observed electronic transport could be considered as an important indicator, albeit indirect, of the underlying difference reflecting the 1O-ODL and 2O-ODL characters of the high temperature states in these two systems, respectively.

Although the Ir dimers in \cis are strictly speaking equivalent to the Ti dimers in \mto, the mechanism of their local formation from the ODL state is electron-hole symmetric and in that sense the dimers in these two systems could be considered as having a different flavor derived from their origin. Formation of dimers in \cis requires transfer of holes from one half of the available population of ODL states to the other, and the ODLs receptors of a hole become dimers, accounting for only 50~\%\ of Ir being dimerized.
In contrast, the dimers in \mto assemble from the ODL states by a virtue of electron transfer, where 1O-ODL states that receive electrons become dimers with all Ti participating in dimerization. While in both systems the process involves bond charge disproportionation, in \cis this consequentially results in the observed site charge disproportionation and subsequent charge order, which is presumably imposed by the specifics of the filling and reflected in the dimer density per formula unit.

\subsection{Consequences for ice-type nanoscale fluctuations}

Presence of 2O-ODL state has another important consequence for \mto. Each Ti$_{4}$ tetrahedron inevitably hosts two 1O-ODL states. Due to the Coulomb repulsion of the bond charge we would expect the two states to be placed on the opposite skew edges of each tetrahedron, although other constellations cannot be excluded. On the other hand, and irrespective of the details of their distribution, multiple 1O-ODLs on one Ti$_{4}$ tetrahedron would cause distortions that are incompatible with the ice-type structural fluctuations in the \mto system such as those suggested in the previous study of the local structure of \mto~\cite{TorigoeNanoscaleicetypestructural2018}. Given that the spin singlet dimer distortion in the ground state of \mto~\cite{SchmidtSpinSingletFormation2004} follows the ``two in -- two out" ice rules~\cite{ander;pr56} on each individual Ti$_{4}$ tetrahedron in the structure, the proposition that the local Ti atomic displacements have the same configuration in the cubic phase~\cite{TorigoeNanoscaleicetypestructural2018} presumably originates in part from the order-disorder type view of the MIT in \mto that would be implicated by the survival of dimer-like distortions in the high temperature phase. Our analysis does not support this picture. However, based on the considerations described above, \cis may possibly be a better candidate for exhibiting the distortions of the ``two in -- two out" type in the disordered ODL regime, as illustrated by mapping shown in Fig.~\ref{fig;cis_odl_review}(b). Exploring this matter further both experimentally and theoretically should provide a more detailed understanding of these systems. Single crystal diffuse scattering based methods~\cite{Weberthreedimensionalpairdistribution2012b,KrogstadReciprocalspaceimaging2020,DavenportFragile3DOrder2019,RothModelfreereconstructionmagnetic2018b,RothSolvingdisorderedstructure2019b} and dynamical mean-field and advanced first-principles approaches~\cite{PramudyaNearlyfrozenCoulomb2011,MahmoudianGlassyDynamicsGeometrically2015,WangTetragonalFeSePolymorphous2020} would be particularly useful in that regard.

\section{Conclusion}
Here we applied joint x-ray and neutron pair distribution function analysis on the dimerized \mto spinel, a candidate system for hosting multi-orbital orbital-degeneracy-lifted (ODL) state, to track the evolution of its local atomic structure across its localized-to-itinerant electronic transition.
Consistent with recent reports, the local structure does not agree with the average structure above the MIT temperature of 250~K and deep in the metallic cubic regime.
However, in stark contrast to previous findings~\cite{TorigoeNanoscaleicetypestructural2018}, we provide unambiguous evidence that spin singlet dimers are vanishing at the MIT.
The shortest Ti-Ti distance corresponding to spin singlet dimers experiences a discontinuous elongation locally on warming through the MIT but remains shorter than that prescribed by the cubic average structure.
The local distortion in the metallic regime is quantitatively and qualitatively different than that observed in association with the spin singlet state, implying that MIT is not a trivial order-disorder type transition.
The distortion characterizes the entire metallic regime, and persists up to at least 500~K ($\sim$2T$_{s}$).
The observed behavior is a fingerprint of the local symmetry broken ODL state observed in the related \cis system.
The correlation length of local distortions associated with the ODL state in \mto is about 1~nm, which is double that seen in \cis, implying two-orbital character of the ODL state.
The observations exemplify that high temperature electronic precursor states that govern emergent complex low temperature behaviors in quantum materials can indeed have a multi-orbital degeneracy lifting character.

\begin{acknowledgements}
Work at Brookhaven National Laboratory was supported by the U.S. Department of Energy (US DOE), Office of Science, Office of Basic Energy Sciences under contract DE-SC0012704. LY and MGT acknowledge support from the ORNL Graduate Opportunity (GO) program, which was funded by the Neutron Science Directorate, with support from the Scientific User Facilities Division, Office of Basic Energy Science, US DOE. Work in the Materials Science Division at Argonne National Laboratory (sample synthesis and characterization) was sponsored by the US DOE Office of Science, Basic Energy Sciences, Materials Science and Engineering Division. X-ray PDF measurements were conducted at 28-ID-1 and 28-ID-2 beamlines of the National Synchrotron Light Source II, a US DOE Office of Science User Facility operated for the DOE Office of Science by Brookhaven National Laboratory.
Neutron diffraction experiments were carried out at the NOMAD beamline of the Spallation Neutron Source, Oak Ridge National Laboratory, which was sponsored by the Scientific User Facilities Division, Office of Basic Energy Science, US DOE.
\end{acknowledgements}

\appendix
\section{Supplemental Results \label{sec_supplementalResults}}



Structural parameters obtained from Rietveld refinements of the tetragonal and cubic models fit to 100~K and 500~K neutron time-of-flight powder diffraction data, respectively, are summarized in Table~\ref{tab;fit_rietveld}.
These confirm the expected average structure behavior of our sample and are in good quantitative agreement with the values reported previously~\cite{SchmidtSpinSingletFormation2004}.

\begin{figure}
\begin{centering}
\includegraphics[width=1\columnwidth]{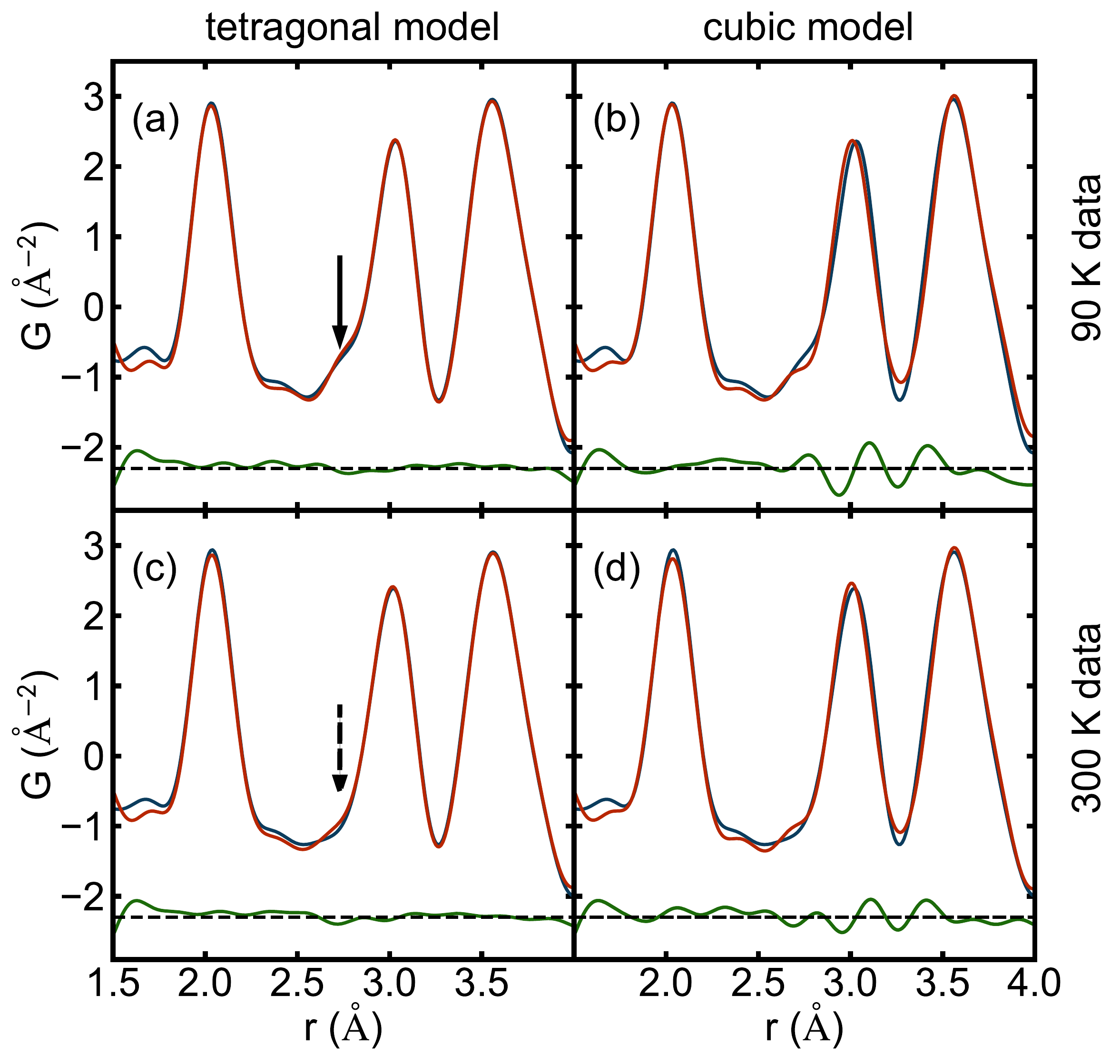}\\
\end{centering}
\caption{(a-d) The x-ray PDF data (blue) collected at the XPD beamline at (a,b) 90~K and (c,d) 300~K fit by (a,c) tetragonal and (b,d) cubic models (red) over the range of $1.5<r<4$~\AA. The difference curves (green) are shown offset below.
}
\label{fig;xray_tetra_cubic_90K_300K_0_4A_fit}
\end{figure}

The data collected in a separate x-ray PDF experiment carried out at a different beamline under the experimental conditions similar to those of the main x-ray PDF experiment are shown in Fig~\ref{fig;xray_tetra_cubic_90K_300K_0_4A_fit}.
The PDF analysis of these data reproduces the observations from the main experiment presented in Fig~\ref{fig;xray_tetra_cubic_90K_300K_0_4A_fit_may}.

\begin{table}
\caption[]{
The Rietveld refinement results of tetragonal and cubic \mto models fit to the neutron TOF powder diffraction patterns at 100~K and 500~K, respectively. The models are introduced in detail in Section~\ref{sec;methods}.
Here, $R_{wp}$ is the weighted profile agreement factor; $x$, $y$ and $z$ are the refinable atomic positions in fractional coordinates; $U_{iso}$, in units of (\AA$^2$), is the isotropic atomic displacement parameter (ADP).
}
\label{tab;fit_rietveld}
\begin{tabular}{lS[table-format=1.6]SSS}
\hline\hline
Temperature (K) & \multicolumn{1}{c}{100}  & \multicolumn{1}{c}{500} \\
Model & \multicolumn{1}{c}{tetragonal}   & \multicolumn{1}{c}{cubic} \\
\hline
$R_{wp}$ 	 & 0.06534 	  & 0.05154    				    \\
$a$ (\AA) 	 & 6.01197 	  & 8.50833	  				\\
$c$ (\AA) 	 & 8.46956	  & \dash   						\\
$c/\sqrt{2}a$& 	0.99616	  & 1.0   						\\
Mg $x$ 	 	 & 0.7449	  & \dash	   		    			\\
Ti ($x$,$y$,$z$) 	 & \multicolumn{1}{c}{(-0.0112,0.2451,-0.1414)}	  & \dash   			\\
O1 ($x$,$y$,$z$)  &   \multicolumn{1}{c}{(0.4795,0.2480,0.1182)}	  &   \multicolumn{1}{c}{(0.2595,0.2595,0.2595)}  					\\
O2 ($x$,$y$,$z$) 	 &  \multicolumn{1}{c}{(0.2410,0.0219,0.8805)} 	  &   \dash  	    \\
Mg $U_{iso}$ (\AA$^2$) & 0.0063  & 0.0110	  					\\
Ti $U_{iso}$ (\AA$^2$) & 0.0093  & 0.0191	   						\\
O $U_{iso}$ (\AA$^2$)  & 0.0042  & 0.0071	   						\\
\hline\hline
\end{tabular}
\end{table}
%

%


\end{document}